\documentclass[twocolumn,floatfix,superscriptaddress,a4paper,showpacs,showkeys,nofootinbib,reprint,prc]{revtex4-1}
\usepackage{epsfig}
\usepackage{latexsym}
\usepackage[colorlinks=true,linktocpage=true,linkcolor=blue,citecolor=blue,allcolors=blue]{hyperref}
\usepackage{url}
\usepackage[utf8]{inputenc}
\usepackage{enumerate}
\usepackage{color}
\usepackage{xcolor}
\usepackage{amsmath,amssymb}

\usepackage[english]{babel}
\usepackage{hyperref}
\usepackage{url}
\usepackage{needspace}
\hypersetup{
   colorlinks=true, %set true if you want colored links
  linktoc=all,     %set to all if you want both sections and subsections linked
  linkcolor=blue,  %choose some color if you want links to stand out
}

\def\be{\begin{equation}}
\def\ee{\end{equation}}
\def\bearr{\begin{eqnarray}}
\def\eearr{\end{eqnarray}}

\def\bfm#1{\mbox{\boldmath $#1$}}

\begin{document}

\title{Momentum dependent potentials from a parity doubling CMF model in UrQMD:\\ Results on flow and particle production}

\author{Jan Steinheimer}
\affiliation{GSI Helmholtzzentrum f\"ur Schwerionenforschung GmbH, Planckstr. 1, D-64291 Darmstadt, Germany}
\affiliation{Frankfurt Institute for Advanced Studies (FIAS), Ruth-Moufang-Str. 1, D-60438 Frankfurt am Main, Germany}

\author{Tom Reichert}
\affiliation{Institut f\"{u}r Theoretische Physik, Goethe-Universit\"{a}t Frankfurt, Max-von-Laue-Str. 1, D-60438 Frankfurt am Main, Germany}
\affiliation{Frankfurt Institute for Advanced Studies (FIAS), Ruth-Moufang-Str. 1, D-60438 Frankfurt am Main, Germany}
\affiliation{Helmholtz Research Academy Hesse for FAIR (HFHF), GSI Helmholtzzentrum f\"ur Schwerionenforschung GmbH, Campus Frankfurt, Max-von-Laue-Str. 12, 60438 Frankfurt am Main, Germany}

\author{Yasushi Nara}
\affiliation{Akita International University, Yuwa, Akita-city 010-1292, Japan}

\author{Marcus Bleicher}
\affiliation{Institut f\"{u}r Theoretische Physik, Goethe-Universit\"{a}t Frankfurt, Max-von-Laue-Str. 1, D-60438 Frankfurt am Main, Germany}
\affiliation{GSI Helmholtzzentrum f\"ur Schwerionenforschung GmbH, Planckstr. 1, D-64291 Darmstadt, Germany}
\affiliation{Helmholtz Research Academy Hesse for FAIR (HFHF), GSI Helmholtzzentrum f\"ur Schwerionenforschung GmbH, Campus Frankfurt, Max-von-Laue-Str. 12, 60438 Frankfurt am Main, Germany}

\date{\today}

\begin{abstract}
The quantum molecular dynamics (QMD) part of the UrQMD model is extended to allow implementation of momentum dependent potentials from a parity doubling chiral mean field (CMF) model. Important aspects like energy conservation and effects on particle production and flow are discussed. It is shown, that this new implementation reproduces qualitatively and quantitatively available data over a wide range of beam energies and improves the description of observables without exception. In particular the description of hyperon and pion production at SIS18 energies is improved. From a comparison with HADES data one could conclude that the present parametrization of the CMF model leads to a slightly too weak momentum dependence. However, a more firm conclusion will require a systematic comparison with flow and multiplicity data over a range of beam energies and system sizes. Our work serves as an important step towards such future studies where the properties of dense QCD matter, through parameters of the CMF model, can be constraint using a comparison of the UrQMD model with high precision heavy ion data, finally also allowing direct comparisons with neutron star and neutron star merger observables.
\end{abstract}

\maketitle

\section{Introduction}

The equation of state (EoS) of dense QCD matter is subject of experimental studies at relativistic heavy ion collider experiments and is important for the understanding of the structure of neutron stars as well as their mergers. While the zero temperature QCD EoS in neutron stars can be inferred from simultaneous measurements of their mass and radius \cite{Miller:2019cac,Riley:2019yda,Miller:2021qha,Riley:2021pdl}, the gravitational waves from binary neutron star mergers \cite{LIGOScientific:2018cki,LIGOScientific:2020aai,LIGOScientific:2020zkf} can offer insights into the EoS at high density and finite temperatures up to 50 MeV \cite{Bauswein:2012ya,Most:2018eaw,Most:2022wgo,Jakobus:2023fru}. 

These astrophysical observations provide complementary information on the properties of QCD matter which have been studied in heavy ion experiments for the past decades at large accelerator facilities \cite{NA49:1999myq,STAR:2002eio,PHENIX:2003nhg,Gazdzicki:2008kk,ALICE:2008ngc,HADES:2009aat}. Here, the focus lies on systematic measurements of particle properties like multiplicities, collective flows as well as momentum correlations which can all be sensitive probes of the underlying equation of state \cite{Stoecker:1980vf,Hartnack:1994ce,Scherer:1999qq,Danielewicz:2002pu,Lacey:2006bc,Hartnack:2011cn,Hartnack:2005tr,STAR:2005gfr,Stoecker:2004qu,Nara:2018ijw}.

At the highest beam energies, e.g. at the top RHIC energy and the LHC, the simulations of such collisions rely on fluid dynamic models in which the equation of state can be easily implemented and studied. Here, lattice QCD calculations predict a smooth chiral crossover, starting at approximately $T=150$ MeV \cite{Borsanyi:2010bp,Bazavov:2011nk,Bazavov:2017dus} which was confirmed by a Bayesian analysis using such fluid simulations and comparing them with experimental data \cite{Pratt:2015zsa,Bernhard:2016tnd}. 

The high density equation of state cannot be calculated directly on the lattice due to the well known sign problem \cite{Fodor:2002km,Allton:2002zi,deForcrand:2002hgr,DElia:2002tig}. In this region different scenarios, including a continuous crossover or even a first-order chiral phase transition are still possible. 

The properties of high density QCD matter can be studied with heavy ion collisions at lower beam energies. These correspond to experiments available at the RHIC-beam energy scan (RHIC-BES II), the SIS18 and upcoming SIS100 at GSI/FAIR as well as at even lower beam energy facilities such as FRIB \cite{Sorensen:2023zkk}. 

From previous studies it is known that for such low collision energies, the fluid dynamic models may not provide meaningful results. The main reason is that the initial non-equilibrium interpenetration phase has a significant impact on final state observables and that already the compression phase depends strongly on the equation of state. Therefore one has to rely on microscopic transport model simulations which are applicable also for non-equilibrium dynamics. A multitude of such models exist and can roughly be divided in two categories, QMD-type models and BUU-type models. The advantages and disadvantages of these models have been discussed extensively in the literature \cite{Aichelin:1989pc,TMEP:2021ljz,TMEP:2016tup}.

Recently, first attempts have been made to constrain the high density behavior of the EoS with such models with statistical inference based on experimental data \cite{OmanaKuttan:2022aml,Oliinychenko:2022uvy}. First results suggest a rather stiff EoS at densities below four times the nuclear saturation density. On the other hand older studies have always highlighted the importance of a momentum dependence of the baryon potentials (used in these simulations) \cite{Aichelin:1987ti,Hong:2013yva,Hartnack:2005tr,Welke:1988zz,Gale:1989dm} and it was recently discussed that a relativistic QMD description will naturally lead to such a momentum dependence \cite{Nara:2020ztb}.

The aim of this paper is to introduce a new way of implementing a momentum dependent potential based consistently on a chiral mean field model for the QCD EoS. In this way, one is not only able to provide a consistent description of the density and momentum dependence of nucleon interactions, but can also include baryon specific potentials and effects of chiral symmetry restoration for all baryon species and their resonances implemented in the transport description. This method will allow, in future studies, to make a statistical inference not only of the density dependence, but also on the momentum and particle type dependence of the interactions leading to a self consistent inference of the EoS. 

In the following we first introduce the chiral mean field model (CMF) which provides the underlying interactions. Then, the implementation of these interactions in the UrQMD transport model is presented and finally we will show the impact of the momentum dependence on flow observables as well as multiplicities measured with the HADES detector at the SIS18 accelerator at GSI.

\section{The CMF model and the EoS}

A consistent equation of state for QCD matter has to fulfill the following constraints: I) It should allow for a description of the data available from heavy ion collisions, II) it should be compatible with all constraints  known from astrophysical observations, and III) it should include all knowledge currently available from lattice QCD simulations. In addition, the number of physical parameters should be limited and meaningful to allow for a satisfactory  physical interpretation.

In addition, one wants to avoid the downsides of purely parametric expansions of the equation of state and its momentum dependence (for example Skyrme density functionals or piece-wise poly-tropes) as such approaches may lead to superluminal speeds of sound, tend to wash-over
structure in the EoS and may be difficult to interpret microscopically. As there is a large number of effective models for the EoS, developed over the last decades, we also want to make sure that our input satisfies as much of known QCD phenomenology as possible. 

For this purpose we employ the fully relativistic parity-doublet Chiral Mean Field (CMF) model \cite{Papazoglou:1998vr,Steinheimer:2010ib,Motornenko:2019arp}. This approach includes, as effective degrees of freedom, a complete set of baryons, interacting with scalar and vector mean field. This includes the full SU(3)-flavor baryonic octet and the $\Delta$ plus their respective parity partners. 

In the hadronic phase, the baryonic octet and all its parity partners interact with the chiral mean-field. In this scenario the effective masses of the ground state octet baryons and their parity partners are dependent on the interactions with the scalar fields $\sigma$ and $\zeta$ and read~\cite{Steinheimer:2011ea}:
\begin{eqnarray}
m^*_{b\pm} = \sqrt{ \left[ (g^{(1)}_{\sigma b} \sigma + g^{(1)}_{\zeta b}  \zeta )^2 + (m_0+n_S m_s)^2 \right]}  \pm g^{(2)}_{\sigma b} \sigma
\end{eqnarray}
Here $+$ stands for positive and $-$ for negative parity states, $g^\text{(j)}_\text{i}$ are the coupling constants of baryons to the two scalar fields, $n_S$ is the strangeness of the baryon so that the SU(3) breaking mass term that generates an explicit mass corresponding to the strangeness. The above mass formula shows a mass splitting between the baryon parity partners which is generated by the scalar mesonic fields $\sigma$ and $\zeta$~\cite{Detar:1988kn,Zschiesche:2006zj,Aarts:2017rrl,Sasaki:2017glk}.

The mean-field values of the chiral fields are determined by the scalar and vector meson interactions, driving the spontaneous breaking of the chiral symmetry. The scalar potential is given as:

\begin{eqnarray}
U_{\rm sc} & = & V_0 - \frac{1}{2} k_0 I_2 + k_1 I_2^2 - k_2 I_4 + k_6 I_6   \nonumber \\
& + & k_4 \ln{\frac{\sigma^2\zeta}{\sigma_0^2\zeta_0}} - U_{\rm sb} ,
\label{veff}
\end{eqnarray}
with
\begin{eqnarray}
    I_2 = (\sigma^2+\zeta^2)&,&~ I_4 = -(\sigma^4/2+\zeta^4),\nonumber\\
    I_6 &=& (\sigma^6 + 4\, \zeta^6)
\end{eqnarray} 
where $V_0$ is included to ensure that the pressure in vacuum  vanishes (i.e. $U_{sc}=0$ for $T=0$ and $\mu_B=0$). The terms $I_n$ correspond to the basic building blocks of possible chiral invariants that form different meson-meson interactions. The logarithmic term in equation (\ref{veff}) introduced in Refs.~\cite{Heide:1993yz,Papazoglou:1996hf}, contributes to the QCD trace anomaly and is motivated by the form of the QCD beta function at the one-loop level. In addition, an explicit symmetry-breaking term is introduced in the scalar potential:

\begin{equation}
U_{\rm sb} = m_\pi^2 f_\pi\sigma +\left(\sqrt{2}m_K^ 2f_K-\frac{1}{\sqrt{2}}m_\pi^ 2 f_\pi\right)\zeta\,.
\label{vsb}
\end{equation}

The vector potential is mediated by the fields: $\omega$ for repulsion at finite baryon densities, the $\rho$ for repulsion at finite isospin densities, and the $\phi$ for repulsion when finite strangeness density is present. The vector fields depend on the respective conserved charge densities and are controlled by the potential $U_{\rm vec}$,
\begin{eqnarray}
U_{\rm vec}&=& -\frac12\left(m_\omega^2\omega^2 + m_\rho^2\rho^2 + m_\phi^2\phi^2\right)\nonumber\\ &-&g_4\left(\omega^4+ \rho^4+\frac12\phi^4 \right)\,.
\end{eqnarray}

The quark degrees of freedom are introduced as in the Polyakov-loop-extended Nambu Jona-Lasinio (PNJL) model~\cite{Fukushima:2003fw}, where their thermal contribution is directly coupled to the Polyakov Loop order parameter $\Phi$ \cite{Motornenko:2019arp}, the quark thermal contribution reads as:
\begin{eqnarray}
	\Omega_{\rm q}=&-&VT \sum_{q_{i}\in Q}\frac{d_{q_{i}}}{(2 \pi)^3}\int{d^3k} \frac{1}{N_c}\ln\left(1+3\Phi e^{-\left(E_{q_{i}}^*-\mu^*_{q_{i}}\right)/T}\right.\nonumber\\
	&+&\left.3\bar{\Phi}e^{-2\left(E_{q_{i}}^*-\mu^*_{q_{i}}\right)/T} +e^{-3\left(E_{q_{i}}^*-\mu^*_{q_{i}}\right)/T}\right)\,,
	\label{eq:q}
\end{eqnarray}
where the index ${q_{i}}$ runs through $u,d,s$ flavors. The antiquark contribution can be obtained by replacing $\mu^*_{q_{i}}\rightarrow-\mu^*_{q_{i}}$, and $\Phi\leftrightarrow\bar{\Phi}$. The Polyakov-loop order parameter $\Phi$ effectively describes the gluon contribution to the thermodynamic potential and is controlled by the temperature dependent potential~\cite{Motornenko:2019arp}:

\begin{eqnarray}
    U_{\rm Pol}(\Phi,\overline{\Phi},T) &=& -\frac12 a(T)\Phi\overline{\Phi} \\ \nonumber
	 + b(T)\ln \Bigl[1-6\Phi\overline{\Phi}\Bigr.
	 &+& \Bigl. 4(\Phi^3+\overline{\Phi}^{3})-3(\Phi\overline{\Phi})^2\Bigr] \,, \\
    a(T) &=& a_0 T^4+a_1 T_0 T^3+a_2 T_0^2 T^2,  \nonumber \\
    b(T) &=& b_3 T_0^4  \nonumber\,.
\end{eqnarray}

The dynamical quark masses $m_q^*$ of the light and strange quarks are also determined by the $\sigma$- and $\zeta$- fields, with the exception of a fixed mass term $m_{0q}$:
\begin{eqnarray}
m_{u,d}^* & =-g_{u,d\sigma}\sigma+\delta m_{u,d} + m_{0u,d}\,,&\nonumber\\
m_{s}^* & =-g_{s\zeta}\zeta+ \delta m_s + m_{0q}\,.&
\end{eqnarray}

Similar to the effective mass $m_b*$ which is modified by the scalar interactions, the vector interactions lead to a modification of the effective chemical potentials for the baryons and their parity partners:

\begin{equation}
    \mu^*_b=\mu_b-g_{\omega b} \omega-g_{\phi b} \phi-g_{\rho b} \rho.
\end{equation}

\begin{figure}[t]
  \centering
  \includegraphics[width=0.5\textwidth]{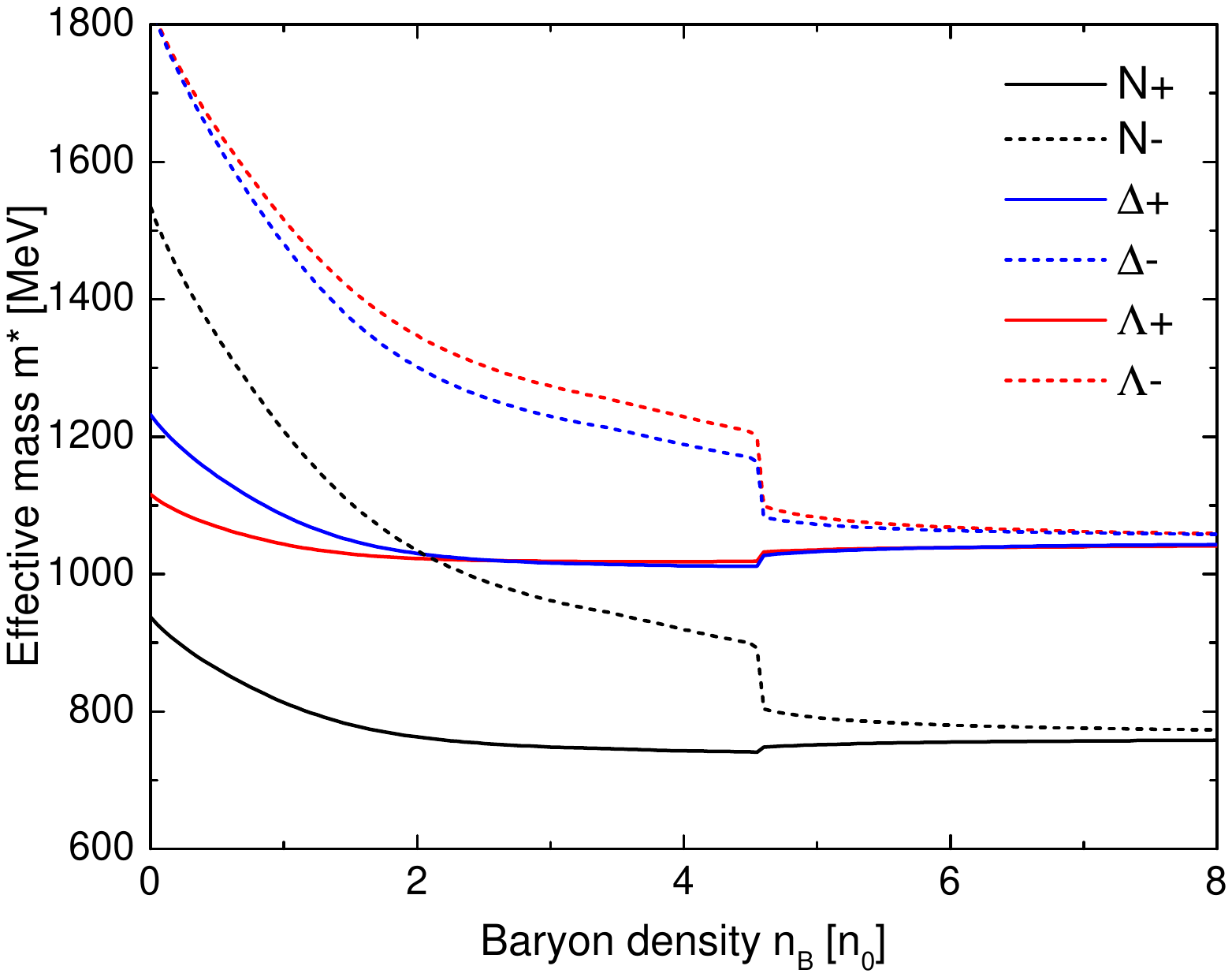}
  \caption{(Color online) Effective masses of baryons and their parity partners for isospin symmetric matter at $T=0$ as function of the baryon density.} 
  \label{fig:0}
\end{figure}

The transition between the quark and hadronic degrees of freedom is controlled by excluded volume interactions, i.e. the suppression of hadrons at high energy densities is maintained by their excluded-volume hard-core interactions~\cite{Rischke:1991ke,Steinheimer:2011ea}.  A volume term introduces an effective chemical potential $\mu^{\rm eff}_j$, which replaces the hadron chemical potential used to calculate the thermal contributions in $\Omega_h$:
\begin{equation}
    \mu^{\rm eff}_j=\mu^*_j - v_j\,P\,,
\end{equation}
for each hadronic particle species $j$. Here, $P$ is the total pressure of the system and the $v_j$ are the EV parameters for the different particle species:

\begin{itemize}
    \item $v_j=0.72~\mathrm{fm}^3$  for baryons;
    \item  $v_j=1/10~\mathrm{fm}^3$ for mesons;
\end{itemize}
while quarks are always assumed point-like.

The parameters of the scalar and the vector interactions and potentials are newly fitted to describe nuclear matter properties at saturation density $n_0=0.16$~fm$^{-3}$ with a binding energy for infinite symmetric nuclear matter as $E/A-m_N =-16.6$~MeV a nuclear incompressibility of $278$ MeV and a symmetry energy of $34$ MeV and symmetry enegy slope of $L=42$ MeV. The hyperon potentials $U_{\Lambda}(n=n_0,p=0)=-30.8$ MeV, $U_{\Sigma}(n=n_0,p=0)=-1.5$ MeV and $U_{\Xi}(n=n_0,p=0)=-16.5$ MeV are fixed by adjusting the corresponding couplings.

In its default version, the CMF model predicts two first-order phase transitions for isospin-symmetric matter. This transition is located at $\mu_B\approx m_N-16$~MeV with critical temperature $T_{\rm CP}\approx 17$ MeV.
At higher densities ($\approx 4.7 n_0$), the CMF model exhibits a first-order phase transition due to the chiral symmetry restoration among baryon parity partners with rather low critical temperature $T_{\rm CP\chi}< 17$ MeV. 
The transition occurs due to the rapid drop in the chiral condensates $\sigma$ and $\zeta$ so the mass gap between parity partners is reduced. This is shown in figure \ref{fig:0} where the effective masses of the nucleons, $\Lambda$ and $\Delta$ baryons are shown, together with their parity partners, as function of the baryon density at $T=0$ for symmetric nuclear matter. As one can see the ground state nucleons mass is reduced only slightly in nuclear matter while the largest effect of chiral symmetry restoration is seen for their parity partners (indicated with a minus-sign) which loose a significant part of their mass. 
 
\begin{figure}[t]
  \centering
  \includegraphics[width=0.5\textwidth]{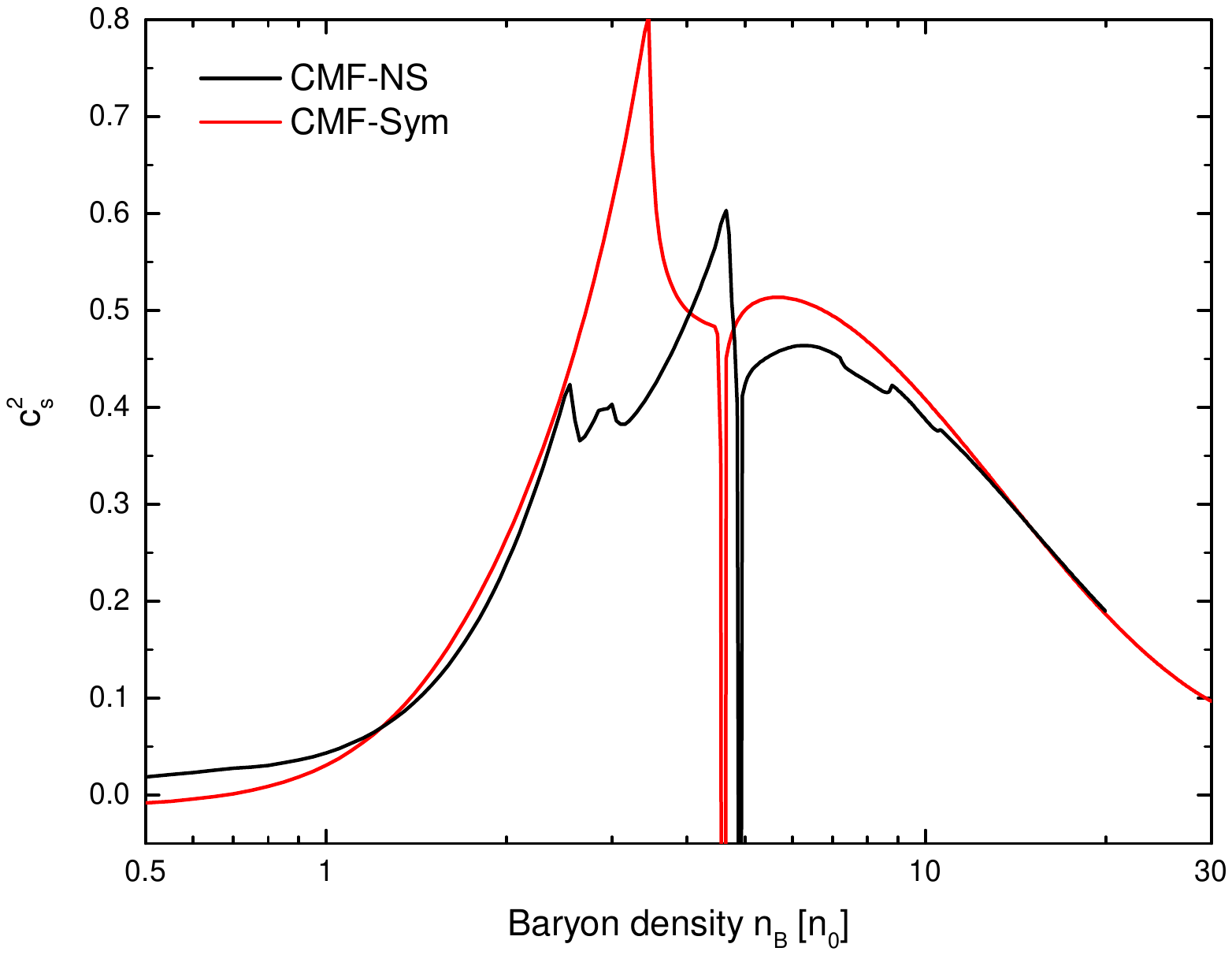}
  \caption{(Color online) Speed of sound at $T=0$ from the CMF model. Shown are the two scenarios for symmetric nuclear matter (red line) and neutron star matter (black line).}
  \label{fig:5}
\end{figure}

Many other aspects of the CMF model have been discussed in previous works (see \cite{Papazoglou:1998vr,Dexheimer:2007tn,Steinheimer:2011ea,Mukherjee:2016nhb,Motornenko:2019arp,Motornenko:2020yme} for the most important aspects) and we will not replicate all of them here. However, let us briefly present the most important features of the present version. Figure \ref{fig:5} shows the speed of sound of dense matter from the CMF model as a function of the baryon density at $T=0$. Two scenarios are compared: The red line depicts the speed of sound for isospin symmetric matter and the black line shows neutron star matter that includes leptons as well as assumes beta-equilibrium, i.e. no strangeness conservation. A general feature of the CMF model is a significant peak in the speed of sound due to the strong hadronic repulsion which is slowly removed as quarks become the dominant degree of freedom.

\begin{figure}[t]
  \centering
  \includegraphics[width=0.5\textwidth]{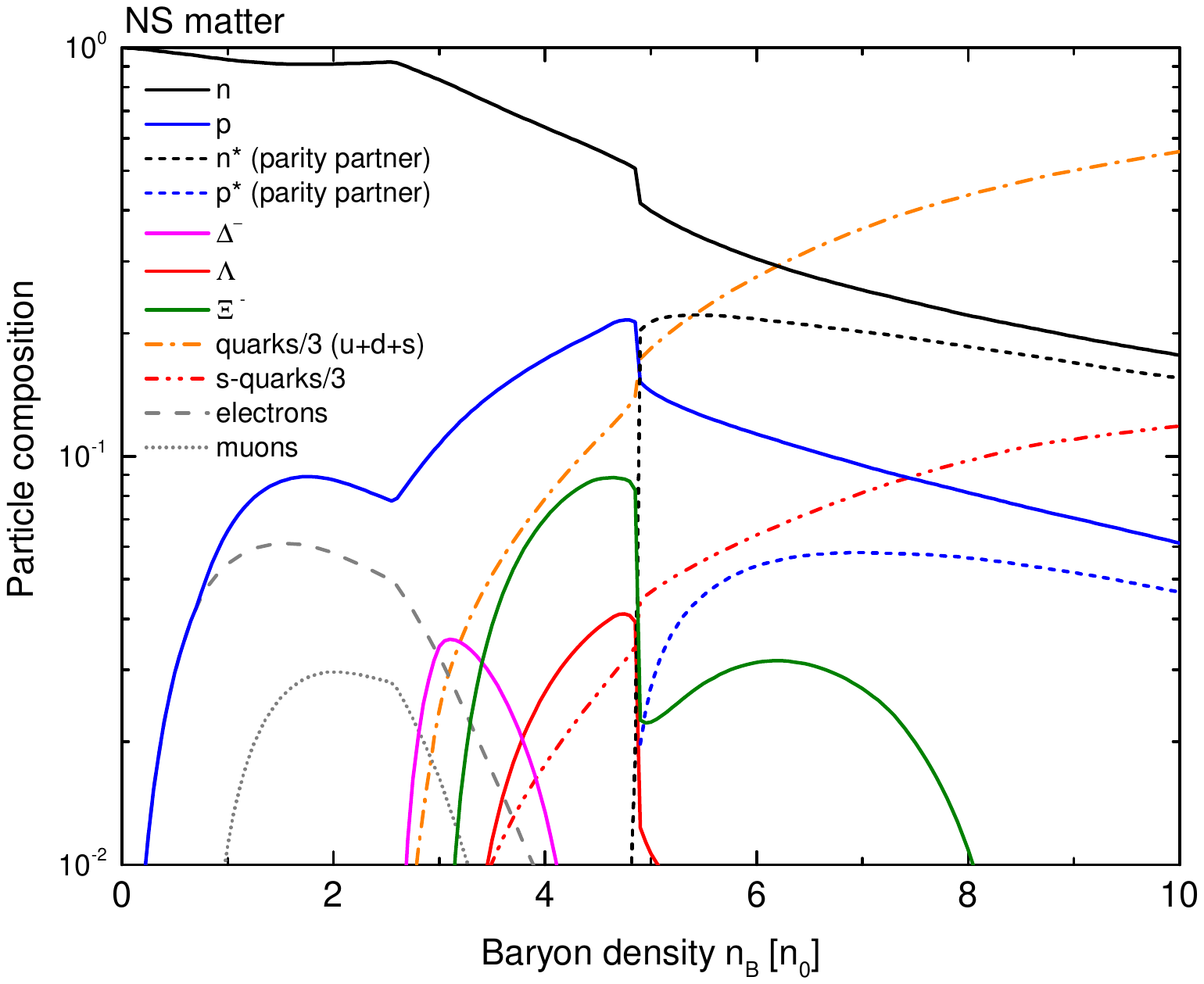}
  \caption{(Color online) Particle composition at $T=0$ for neutron star matter as function of density.}
  \label{fig:6}
\end{figure}

The transition of the degrees of freedom can also be seen in figure \ref{fig:6} which shows the particle composition for neutron star matter as a function of the net baryon density. What is important to note here is that the transition from hadronic to deconfined quark degrees of freedom proceeds smoothly at a density that is beyond 6 times nuclear saturation density. Before that however, the effects of chiral symmetry restoration, i.e. the degeneracy of the parity partners are clearly visible.

The equation of state of the CMF model has been used in UrQMD simulations before \cite{OmanaKuttan:2022the,Steinheimer:2022gqb}. However, these simulations used only the density dependence of the equation of state, i.e. a momentum-averaged EoS. However, previous works have been highlighting the importance of the momentum dependence of the potential interactions \cite{Aichelin:1987ti,Hong:2013yva,Hartnack:2005tr,Welke:1988zz,Gale:1989dm}. Fortunately the fully relativistic nature of the CMF model provides the opportunity to calculate the momentum dependent potentials for a large set of baryons in the UrQMD model and allows us to implement them consistently in a straightforward way.

\subsection{Momentum dependent single particle energies in the CMF}\label{sec:CMF}

To provide the equation of state of the CMF model for the UrQMD simulations, momentum dependent single particle energies $U_i(n_B,p)$ are necessary.

\begin{figure}[t]
  \centering
  \includegraphics[width=0.5\textwidth]{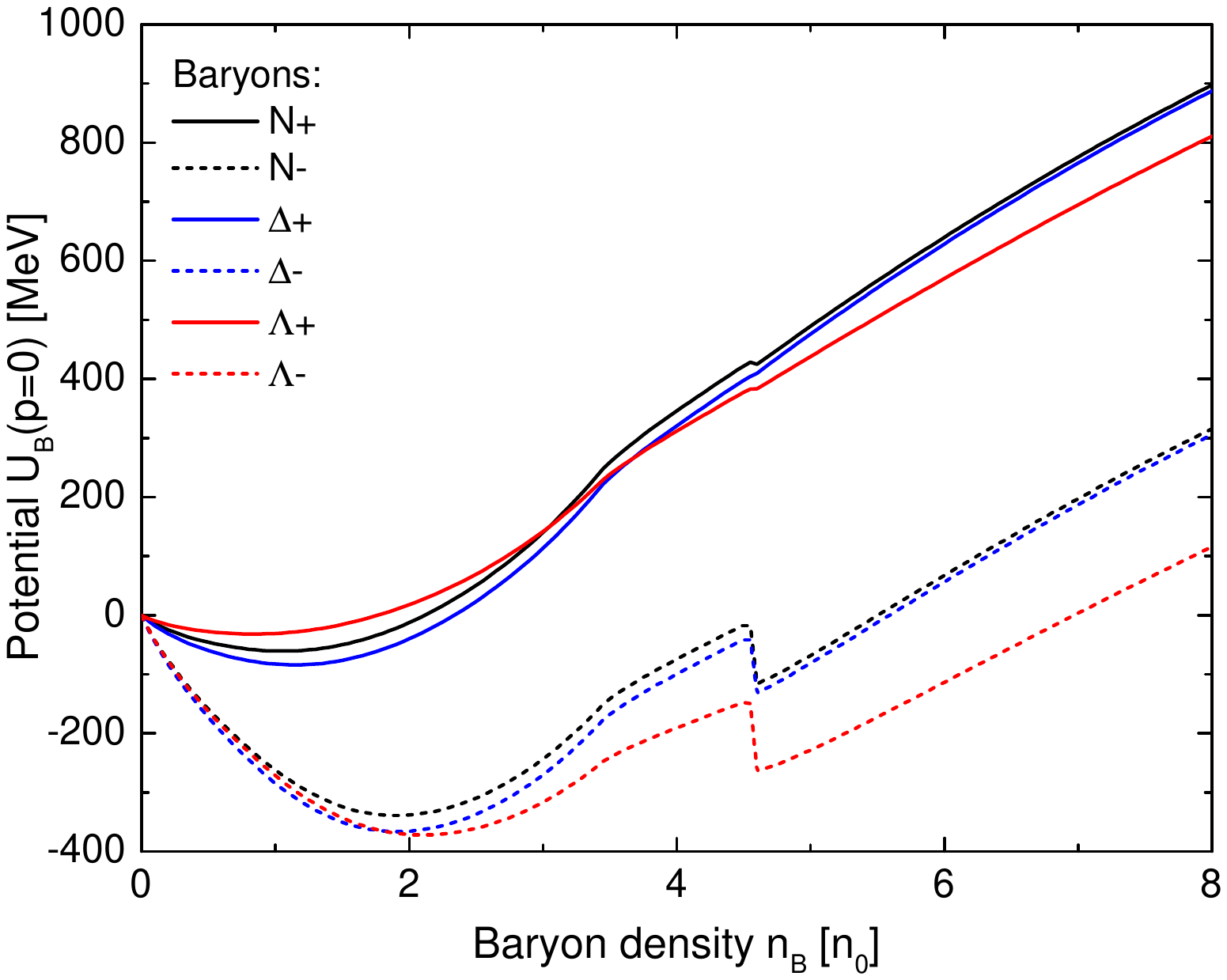}
  \caption{(Color online) Single particle potentials of different baryons and their parity partners at $T=0$ and $n_B = n_0$, for zero momentum.} 
  \label{fig:1}
\end{figure}

The momentum dependent optical potential for a baryon species $i$ can be defined as the difference between the in-medium single particle energy and its corresponding vacuum energy at a given momentum $p$:

\begin{eqnarray}\label{optical}
  U_i(n_B,p) & = & E^*-E_{vac}  \\
  & = & \sqrt{m^{*2}+p^2} - \sqrt{m_0^2 + p^2} -(\mu_i^{eff}-\mu_i) \nonumber
 \end{eqnarray}

where we ignore the role of the vector-Lorenz part. As can be seen, for a vanishing momentum $p$, the optical potential reduces to the difference between the scalar potential and the vector potential ${U_i(p=0)=m^*-m_0-\mu_i^{eff}+\mu_i}$. This optical potential $U_i(p=0)$ is shown, as function of the net baryon density for $T=0$, in figure \ref{fig:1} for nucleons, hyperons and $\Delta$s together with their parity partners. 
As one can see, the optical potential for hyperons, at densities up to 3 times saturation density, is larger than for nucleons which is expected as the binding energy is smaller. The $\Delta$ baryon is more deeply bound, however does not appear in normal nuclear matter due to its larger mass \footnote{Note that if the vector coupling of the $\Delta$ would be larger, then also the binding energy for it will be smaller.}. Interestingly, all the parity partner states are much deeper bound than the ground state baryons due to the larger coupling to the attractive scalar field. This has been discussed as a possible source of additional correlations between parity partners in dense matter \cite{Koch:2023oez} and could in principle be studied within our approach. Also, the effect of the chiral transition appears much larger for the parity partners as their effective mass is reduced significantly at this transition, while the ground state masses remain essentially unchanged.

Going back to equation (\ref{optical}) one can furthermore deduce that for very large momenta, the role of the (effective) mass vanishes and the optical potential reduces to simply the repulsive vector potential giving an asymptotic value of the momentum dependent optical potential of 
\begin{equation}
    U_b(p\rightarrow \infty) = -\mu_i^{eff}+\mu_i =g_{\omega i} \omega + g_{\phi i} \phi + g_{\rho b} \rho.
\end{equation}

\begin{figure}[t]
  \centering
  \includegraphics[width=0.5\textwidth]{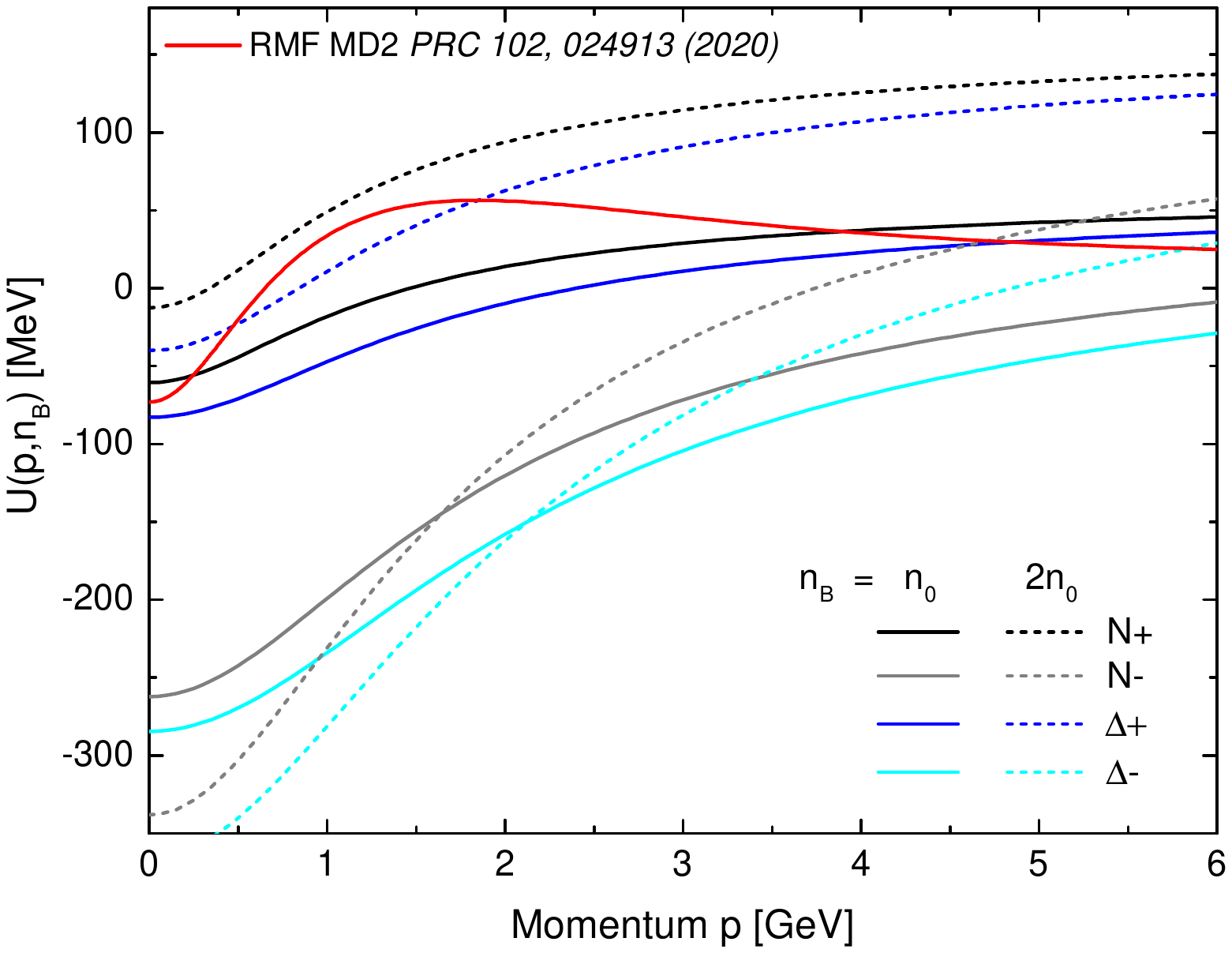}
  \caption{(Color online) Single particle potentials of nucleons and $\Delta$ and their parity partners at $T=0$ as function of momentum for two different densities. As reference the potential from a RMF model, at nuclear saturation density \cite{Nara:2020ztb}, is also shown as red curve. The CMF potential shows a much weaker momentum dependence due to the reduced scalar coupling in the parity doubling realization.} 
  \label{fig:2}
\end{figure}

This is depicted in figure \ref{fig:2} where the momentum dependence of the optical potential of nucleons and $\Delta$ is shown for two different baryon densities $n_B=n_0$ (solid lines) and $n_B=2n_0$ (dashed lines). It is clear that the ground state baryons and their parity partners show clearly distinct density and momentum dependencies. This is clearly different from the conventional approaches of momentum dependent potentials where a single parameterized momentum dependency is often used for all baryon types and densities. In our approach we are able to provide a density and momentum dependence of the optical potential which is consistent with the underlying equation of state.

\begin{figure}[t]
  \centering
  \includegraphics[width=0.5\textwidth]{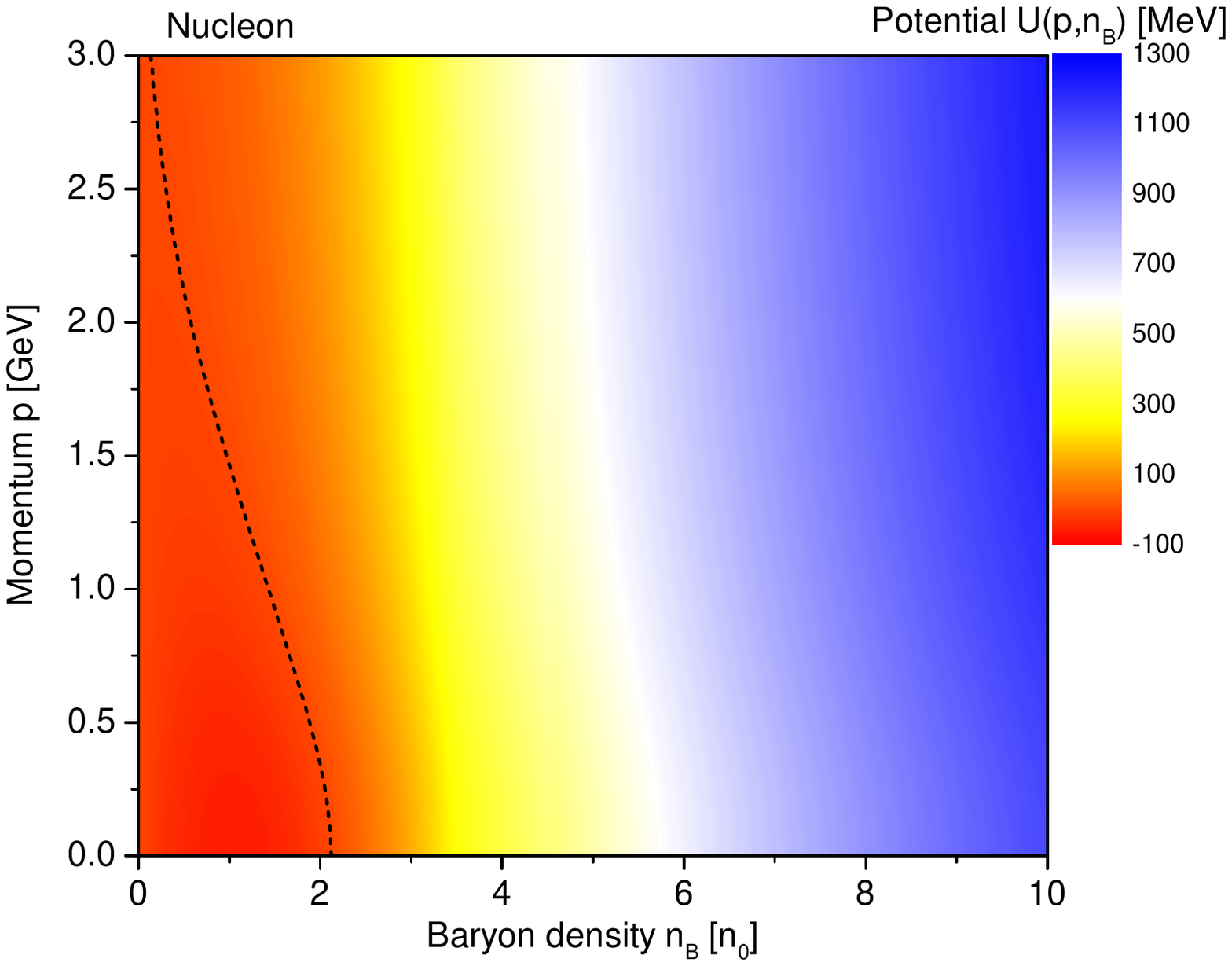}
  \caption{(Color online) Single particle potentials of nucleons at $T=0$ as function of momentum and baryon density. $U(p,n_B)=0$ is shown as dashed contour.} 
  \label{fig:3}
\end{figure}

\begin{figure}[t]
  \centering
  \includegraphics[width=0.5\textwidth]{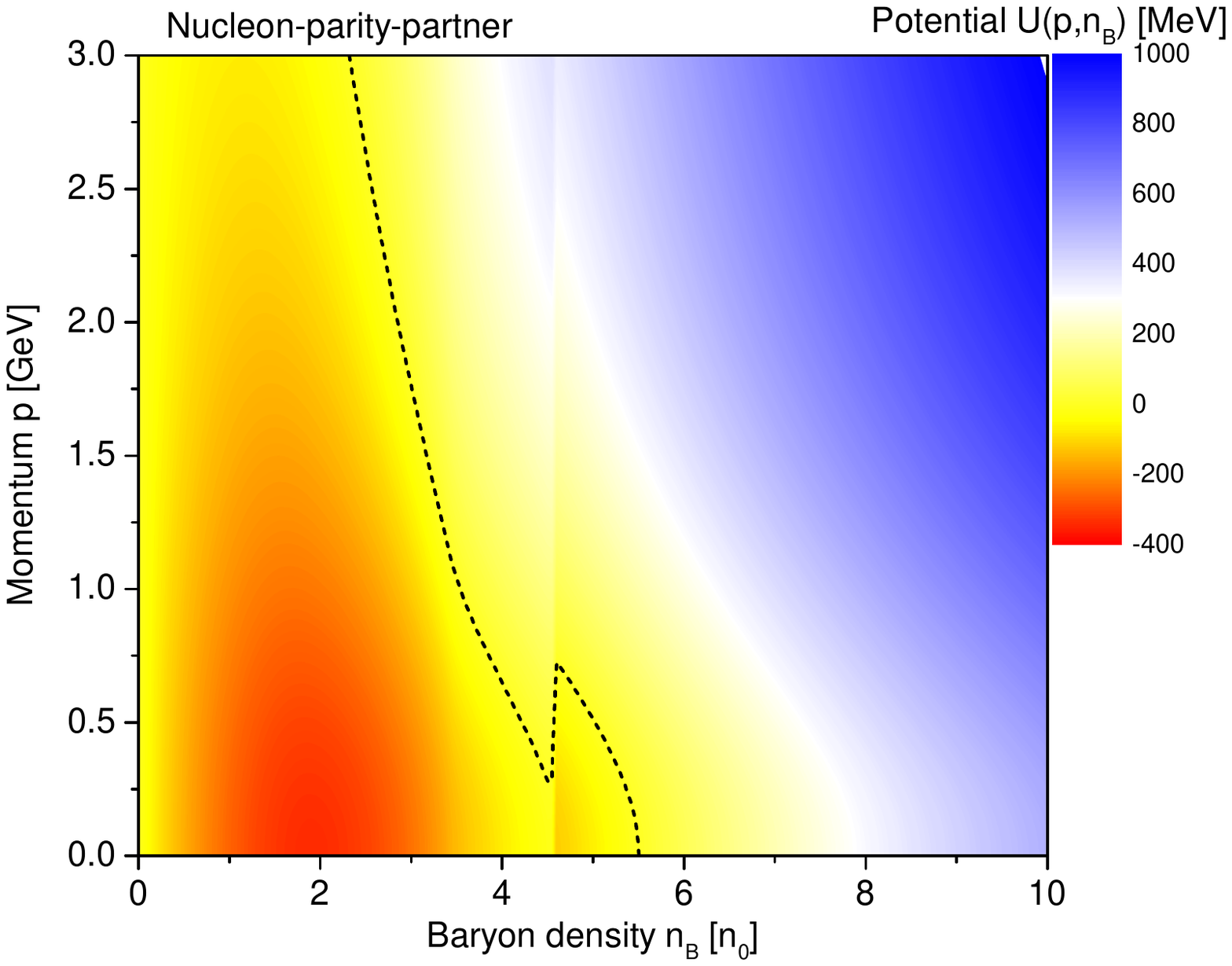}
  \caption{(Color online) Single particle potentials of the parity partner of the nucleons at $T=0$ as function of momentum and baryon density. $U(p,n_B)=0$ is shown as dashed contour.} 
  \label{fig:4}
\end{figure}

For completeness, figures \ref{fig:3} and \ref{fig:4} show the full momentum and density dependence of the optical potential of the nucleon and the nucleon parity partner at $T=0$ and for isospin symmetric nuclear matter. Again, the chiral transition is much more pronounced in the parity partner potential, indicated by the significant jump of the $U=0$ line (dashed line) visible only in the figure for the parity partner.

As a final remark we should also mention that in the present approach we do not assume any explicit momentum dependence of the coupling parameters to the fields which means that the transition occurs at the same density for all momenta. In principle, one may imagine a momentum dependent coupling which e.g. could lead to a vanishing optical potential at infinite momentum. However, solving such a new set of equations self-consistently is outside the scope of the current work.

\section{The UrQMD model}

After defining the momentum and density dependent optical potentials for the whole set of baryons we need to introduce the framework how these can be implemented in the UrQMD approach~\cite{Bass:1998ca,Bleicher:1999xi,Bleicher:2022kcu}. The cascade part of the model is based on the propagation of hadrons on classical trajectories in combination with stochastic binary scatterings, color string formation, and resonance excitation and decays. The imaginary part of hadron interactions are based on a geometric interpretation of their scattering cross sections, which are either taken from experimental measurements where available \cite{ParticleDataGroup:2020ssz}, or are calculated, e.g., from the principle of detailed balance. The real part of hadronic interactions is done with a quantum molecular dynamics (QMD) approach. Up to now, only a density-dependent potential interaction term based on early, non-relativistic QMD approaches~\cite{Aichelin:1991xy} which incorporated density-dependent Skyrme interactions \cite{Hartnack:1997ez} was taken into account in the standard version.

The non-relativistic equations of motion for the QMD model  \cite{Aichelin:1986wa} are given as:

\begin{eqnarray}\label{motion}
\dot{\textbf{r}}_{i}=\frac{\partial \mathrm{\bf{H}}  }{\partial\textbf{p}_{i}},
\quad \dot{\textbf{p}}_{i}=-\frac{\partial \mathrm{\bf{H}} }{\partial \textbf{r}_{i}},
\end{eqnarray}
where $ \mathrm{\bf{H}} = \sum_i H_i$ is the total Hamiltonian function of the system which is the sum over all Hamiltonians, $H_i=E^{\mathrm{kin}}_i + V_i$, of the $i$ baryons.
It includes the kinetic energy and the total potential energy
$ {\mathrm{\bf{V}}=\sum_i V_i \equiv \sum_i V\big(n_B(r_i)\big)}$. The change of momentum of each baryon can be calculated from Hamilton's equations of motion,
\begin{eqnarray}
\dot{\textbf{p}}_{i} & = &-\frac{\partial \mathrm{\bf{H}}}{\partial \textbf{r}_{i}} =  -  \frac{ \partial \mathrm{\bf{V}} }{\partial \textbf{r}_{i}} \\
  & = & - \left(\frac{ \partial V_i }{\partial n_i}\cdot \frac{\partial n_i}{\partial \textbf{r}_{i}} \right)-\left( \sum_{j \ne i} \frac{\partial V_j}{ \partial n_j} \cdot \frac{\partial n_j}{\partial \textbf{r}_{i}}\right) ~,
\label{QMD_eq}
\end{eqnarray}
where $n_{\{i,j\}} \equiv n_B (\bfm{r}_{\{i,j\}})$ is the local interaction density of baryon $i$ or $j$.
Thus, $V_i$ corresponds to the average potential energy of a baryon at position $\bfm{r}_i$, and the local interaction density $n_B$ at position $\bfm{r}_k$ is calculated by assuming that each particle can be treated as a Gaussian wave packet~\cite{Aichelin:1991xy,Bass:1998ca}. With such an assumption, the local interaction baryon density $n_B(\bfm{r}_k)$ at location ${\bf r}_k$ of the $k$-th particle in the computational frame is:
\begin{eqnarray}
    n_B(\bfm{r}_k) &=& n_k = \sum_{j,\,j\neq k} n_{j,k} \\ \nonumber 
    & = & \left(\frac{\alpha}{\pi}\right)^{3/2}\sum_{j,\,j\neq k} B_j \exp{\left(-\alpha({\bf r}_k-{\bf r}_j)^2\right)} \, , \label{gauss_qmd}
\end{eqnarray}
where $\alpha=\frac{1}{2 L}$, with $L=2$ fm$^2$, is the effective range parameter of the interaction. The summation runs over all baryons, and $B_j$ is the baryon charge of the $j$-th baryon. Once the potential energy per baryon is known, the equation (\ref{QMD_eq}) can be solved numerically. Note that a purely density dependent CMF potential has been implemented in UrQMD in a previous work and it was shown that the resulting EoS very closely reproduces the expected CMF results \cite{OmanaKuttan:2022the}. This was achieved by directly extracting the potential energy as 
\begin{equation}
    V_{\mathrm{CMF}} = E_{\mathrm{field}}/A  \equiv E_{\mathrm{CMF}}/A - E_{\mathrm{FFG}}/A\,,
\end{equation}
and not from the single particle potential $U(n_B)$. In this set-up the total momentum is conserved exactly by construction, which is different from the momentum dependent potentials, as we will see below. In the following we will use this purely density dependent method as baseline for our momentum dependent implementation which is based on the single particle potentials.

\subsection{Implementing the CMF-momentum dependence}

The momentum dependence of the optical potential in QMD models can be understood as a way to mimic the appearance of an effective mass $m^*$ due to a scalar interaction which would then in effect modify the kinetic part of the self energy. 

\begin{figure}[t]
  \centering
  \includegraphics[width=0.5\textwidth]{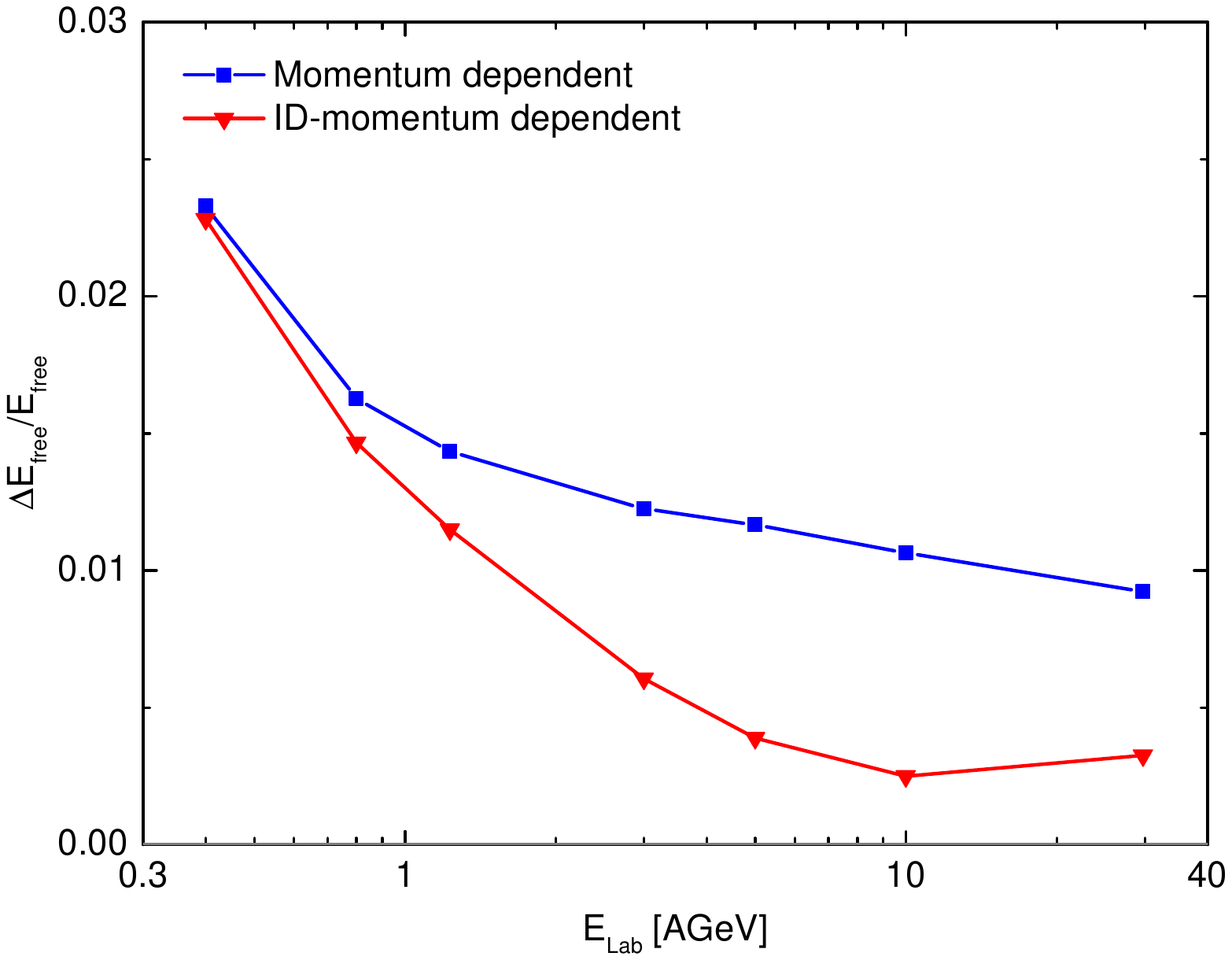}
  \caption{(Color online) Relative difference, in the available energy, using the momentum dependent potentials compared to simulations with only density dependent potentials, which conserve momentum exactly. The available energy $E_{free}$ is defined as the total energy minus $2A \cdot m_N$.}
  \label{fig:7}
\end{figure}

Note that implementing the fully relativistic equations of motion, as e.g. presented in \cite{Nara:2020ztb}, would naturally include the effects of the scalar and vector interactions and such an implementation as discussed here would not be necessary. In its current version the UrQMD model still uses the non-relativistic equations of motion in case of potential interactions, and thus one needs to implement the momentum dependence explicitly. This can be done by calculating the Schr\"odinger equivalent potential from the relativistic theory \cite{Botermans:1990qi,Ainsworth:1987hc}. However, this leads to unrealistic momentum dependencies, for example a linear increase of $U$ with energy (see e.g. \cite{Jaminon:1989wj,Fuchs:2005yn}). This problem is addressed in most cases by simply parameterizing a momentum dependence to fit the one measured from proton nucleus scattering experiments \cite{Hama:1990vr}. In our work we will go a slightly different path and use the momentum dependence from the relativistic CMF single particle energy directly, i.e. a CMF inspired effective parametrization of the momentum dependence which is consistent with the underlying equation of state.

As the CMF model provides the single-particle potential $U_i = U(n_B,p)$ one has to calculate the potential energy $V_i(n_B,p)$ accordingly. For an only density dependent potential this can be done through solving 
\begin{equation}
\label{eq:U}
U(n_B)=\left. \frac{\partial \big(n_B  \cdot V(n_B)\big)}{\partial n_B}  \right|_{p=p_{\mathrm{Fermi}}} 
\end{equation}
at the Fermi momentum. 

To obtain the explicitly momentum dependent $V(n_B,p)$ we numerically invert (and integrate) equation (\ref{eq:U}) for any fixed momentum $p$:

\begin{equation}
\label{eq:V}
V(n_B,p)= \left. \frac{1}{n_B} \int_{0}^{n_B} U(n_B,p) dn_B \right|_{p=\mathrm{const.}} 
\end{equation}

The only input are the single particle energies $U_i(n_B,p)$ from CMF given in eq. (\ref{optical}). For now, we use the optical potential for symmetric nuclear matter and exact strangeness conservation only, thus the values of the strange vector field $\phi$ and iso-spin density dependent vector field $\rho$ are exactly zero, leaving us with only the $\omega$ contribution to the vector potential. This considerably simplifies equation (\ref{optical}) and allows us to calculate $U$ purely as function of the net baryon density $n_B$. Having obtained $V_i(n_B,p)$ one can then numerically calculate $\frac{ \partial V_i }{\partial n_B}$ at a fixed momentum p.

At this point, one has to take into account one caveat. In the CMF model, the momentum $p$ is defined in the rest-frame of the mean-background-field of the medium. This can be directly transferred to mean-field type transport simulations, like BUU models for example, where the relative momentum with respect to some event-averaged external mean field is well defined. In the QMD approach there is no such external background field, as all fields are created by the other particles in the same event at given distance $r_k-r_j$. For all the following we will assume that $\Delta p = p_k-p_j$ which is the difference of the momenta of the two particles $k$ and $j$ in the computational frame. This basically assumes that each particle creates a (background)field for any other particle \footnote{It is interesting to note that for a Fermi distribution this equality is almost exact as one can show that the average distance between two points in a homogeneous sphere of radius $R$, in three dimensions, is almost exactly $R$.}. Such an approximation has also been successfully employed in previous works and the differences from results, assuming a Fermi distributed momentum at T=0 and at finite temperature was discussed \cite{Welke:1988zz}.

As a first test we want to check how well this implementation conserves the total energy. In general, when a momentum dependent potential is employed, the two-body collision term violates energy-momentum conservation as it changes the momenta without respecting the equations of motion (see e.g. \cite{Ikeno:2023cyc} for a way how to improve the momentum conservation). In order to study the severity of the violation of energy conservation, we simulated three different scenarios.

\begin{enumerate}
\item \textit{Density dependent}: These simulations follow our prescription in \cite{Steinheimer:2022gqb}, using a density dependent potential based on our CMF parametrization. This can be considered our baseline.
\item \textit{Momentum dependent}: For these simulations we use our CMF-momentum dependent prescription described in this section. However we employ the momentum and density dependent potentials of ground state nucleons (from CMF) for all baryons. This is a common procedure in many transport simulations, with only a handful of exceptions (see e.g. \cite{Boguta:1982rsg,Waldhauser:1987uk,Hartnack:1993bp,Larionov:2021ycq}.
\item \textit{ID-Momentum dependent}: Here, we employ the full set of momentum dependent potentials from CMF as presented in section \ref{sec:CMF}. This means ground state nucleons, $\Delta$s and hyperons have separate potentials as well as their parity partners. This is the first time such a complete set of ID-dependent potentials is implemented in a transport simulation for heavy-ion reactions.
\end{enumerate}

Figure \ref{fig:7} shows the relative violation of energy conservation in the case with momentum dependent potentials. The quantity shown $E_{free}$ is defined as the total available energy in the collision system, i.e. the total invariant mass of both colliding nuclei minus the sum of the vacuum masses of all colliding nucleons $E_{free}=\sqrt{s_{AA}}- 2 A m_N$. The difference $\Delta E_{free}$ is then defined as the difference with only the density dependent potentials. As one can see the relative violation of the total energy conservation is very small. The relative violation is largest at the lowest beam energies, because here the available energy is smaller than at larger beam energies. Also we can observe that consistency in the potentials, i.e. using the full set of ID-dependent potentials, improves the energy conservation. Overall, the effect of the energy conservation violation is very small ($< 2.4 \% $) for all energies considered.

\begin{figure}[t]
  \centering
  \includegraphics[width=0.5\textwidth]{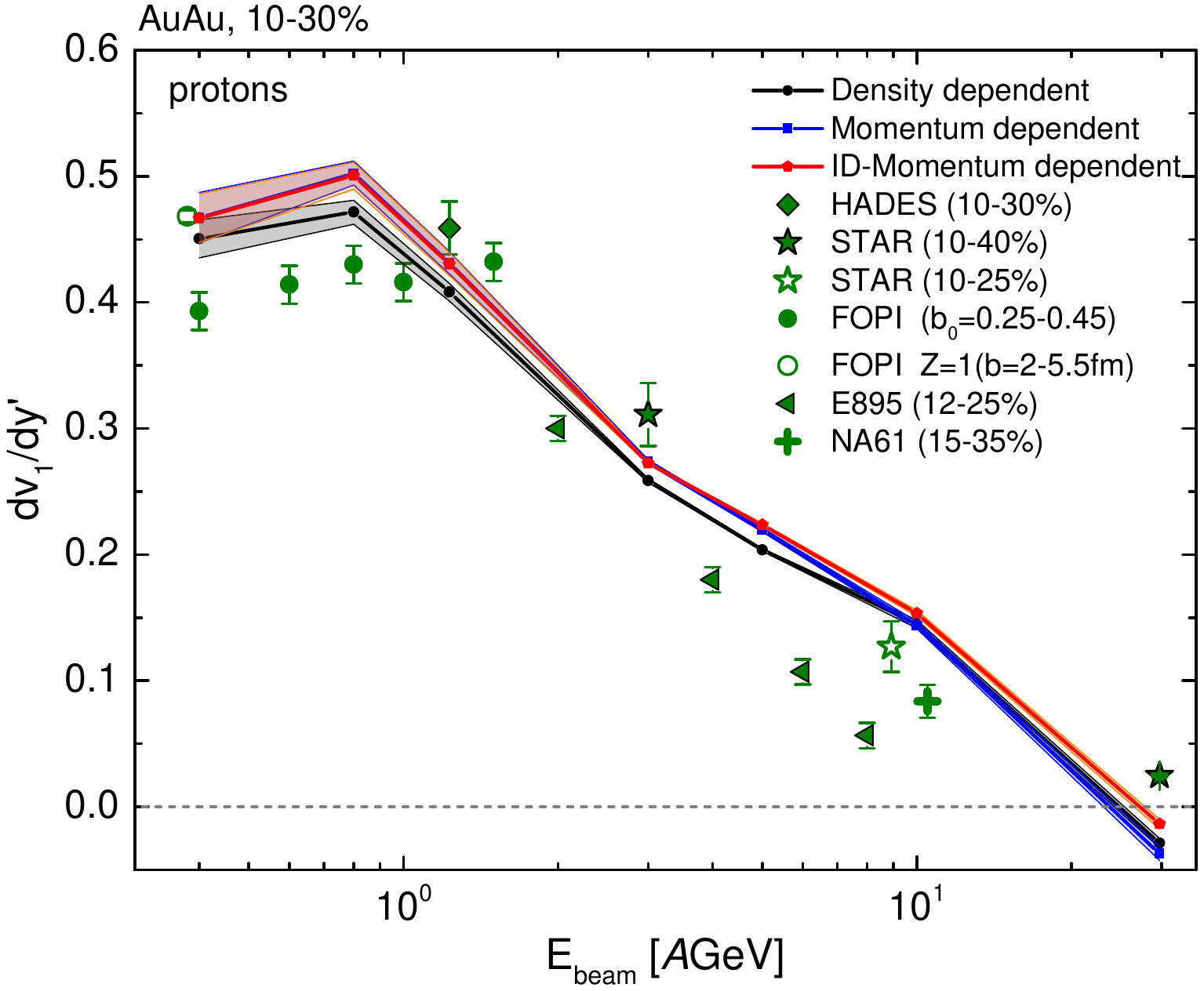}
  \caption{(Color online) Slope of $v_1$ of protons at mid-rapidity as function of beam energy. The three UrQMD scenarios are shown with different colors.}
  \label{fig:8}
\end{figure}

\section{Results}

In the following we present results on the effect of the CMF-momentum dependent potentials on various observables in heavy ion collisions. First a comparison with world data on the directed and elliptic flow will be shown and later we will focus more specifically on data provided by the HADES collaboration which recently presented a high statistics results on differential flow and multiplicity observables in Au+Au collisions at $E_{\mathrm{lab}}= 1.23 A$ GeV.  

Figure \ref{fig:8} shows the mid-rapidity slope of the directed flow of protons for mid-central collisions, simulated with UrQMD and the three scenarios presented above. The usual density dependent potentials are shown as black lines and the momentum dependent cases are presented as red and blue curves. The various data \cite{Andronic:2001sw,Andronic:2004cp,Andronic:2006ra,Andronic:2004cp,FOPI:2011aa,Pinkenburg:1999ya,Liu:2000am,E877:1997zjw,Barrette:1994xr,E877:1996czs,HADES:2022osk} are shown as different green symbols. One should note that the data are only helpful to give a rough guide as the different experiments often use differing centrality definitions and acceptance ranges as well as different methods of extracting the mid-rapidity slope \footnote{We employ a third order polynomial fit to the $v_1(y')$ where $y'$ is the rapidity scaled to the beam rapidity in the center of mass frame.}. What is obvious is that the effect of the momentum dependence is rather small for the integrated directed flow and the difference between the two methods is even smaller. The largest effect can be observed at the lowest beam energies.

\begin{figure}[t]
  \centering
  \includegraphics[width=0.5\textwidth]{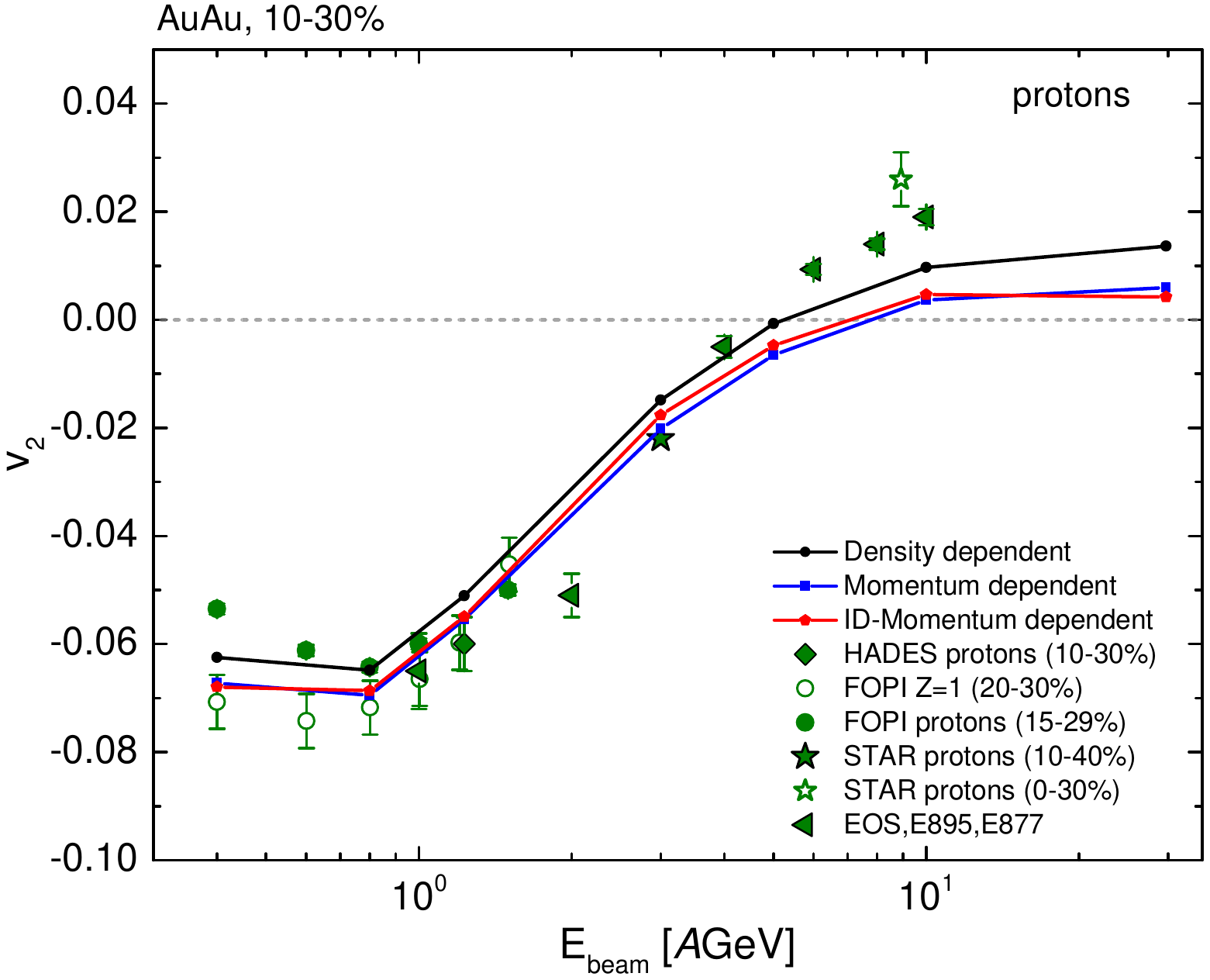}
  \caption{(Color online) Elliptic flow $v_2$ of protons at mid-rapidity as function of beam energy.}
  \label{fig:9}
\end{figure}

A similar conclusion can be drawn for the elliptic flow $v_2$ which is shown in figure \ref{fig:9} in the same colors. The relative difference for the integrated elliptic flow appears a bit larger than for the directed flow, however the effect is still small compared to the uncertainties arising from the different experimental analyses. One thing that is observed for both flow coefficients is that the model becomes less reliable at the higher beam energy, above $\sqrt{s_{NN}}>5$ GeV and appears to be too stiff. This can be understood from the fact that we still employ a non-relativistic QMD approach which will have a maximum compression due to the lack of Lorenz contraction of the Gaussian wave-packages at the highest beam energies.
In addition, it is important to mention that the current version of the CMF model contains deconfined degrees of freedom which are not part of the UrQMD transport model. This means that whenever the system reaches densities at which the quarks significantly contribute to the EoS, the effective EoS introduced by the potentials will not be consistent with the CMF mean field equation of state. Such densities are only reached by the highest beam energies in this study. 
This means that the effective EoS will never be probed much beyond its stiffest point and therefore appear stiffer as it would be at these energies. Improving this situation would require implementing relativistic corrections or even a fully covariant description \cite{Sorge:1989dy,Nara:2019qfd} which is out of the scope of the current work. Any results for the highest beam energies should therefore be taken with a grain of salt.

\subsection{Comparison with HADES data}

\begin{figure}[t]
  \centering
  \includegraphics[width=0.5\textwidth]{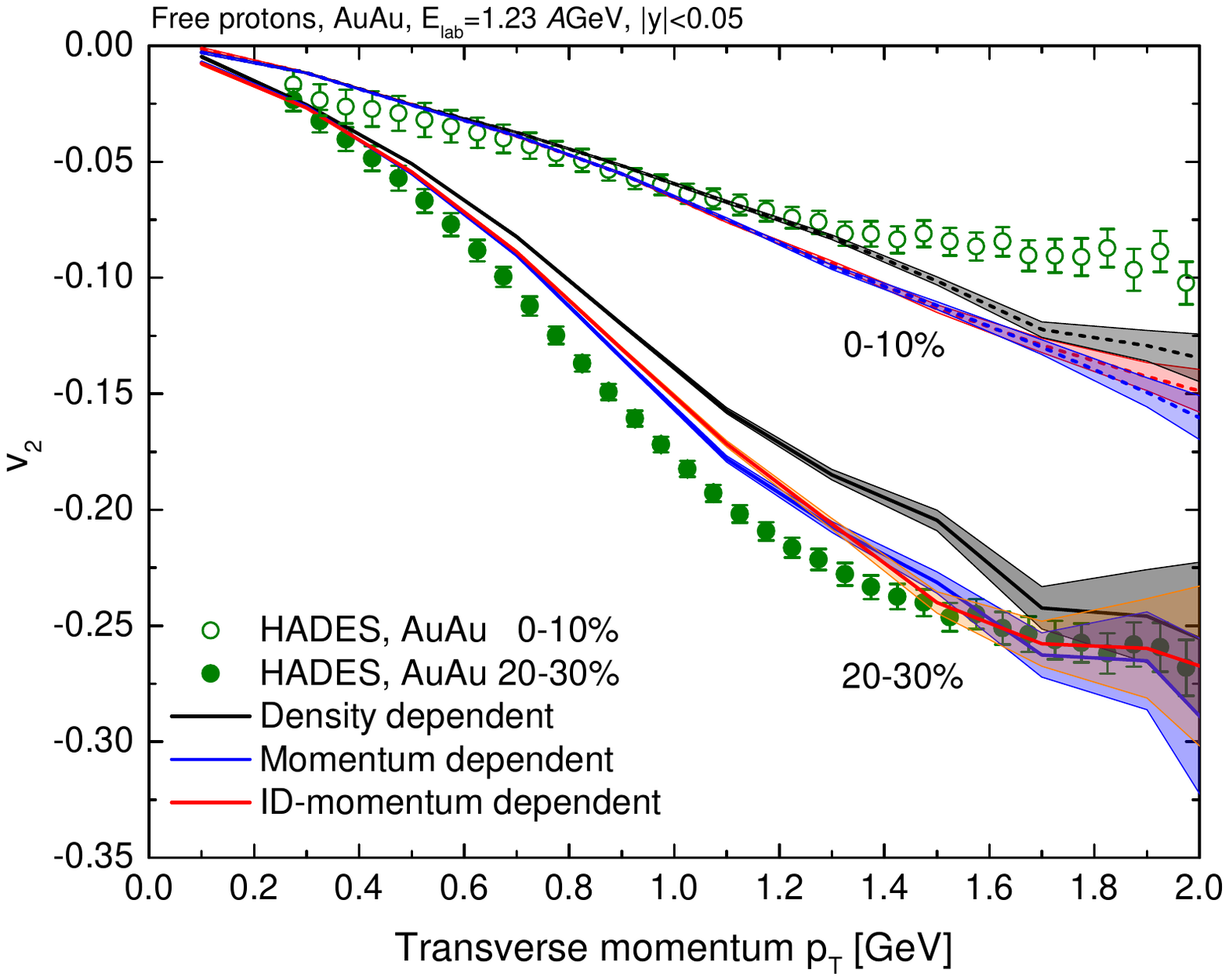}
  \caption{(Color online) Elliptic flow $v_2$ of protons at mid-rapidity as function of transverse momentum for Au+Au collisions at $E_{lab}= 1.23 A$ GeV. Two different centralities are compared, central (dashed lines) and mid-central (solid lines), with data from HADES \cite{HADES:2022osk,HADES:2020lob}.}
  \label{fig:10}
\end{figure}

The effects of the momentum dependence should be better visible in more $p_T$-differential observables. The $p_T$ dependent elliptic flow for example has been measured with high statistics for protons, by the HADES experiment \cite{HADES:2022osk}. Figure \ref{fig:10} shows the momentum dependence of the elliptic flow for the three different QMD scenarios compared to HADES data (symbols). Shown are only two different centrality selections for better visibility. For central collisions, the effect of the momentum dependence is negligible, however all three simulations tend to overestimate the elliptic flow at large momenta. For more peripheral collisions (20-30$\%$), the effect of the momentum dependence becomes more visible and including the CMF-momentum dependence improves the description of the data. Again the difference between the two momentum dependent scenarios is not very large. One may argue that the data indicate that the momentum dependence in the present version of CMF is too small, which could be addressed by changing the scalar coupling parameters. A stronger momentum dependence then will generally lead to a slightly larger drop in the effective mass for the ground state nucleon. Such a fine-tuning of the CMF model however is not the purpose of the current work but rather to explore general dependencies and present a functioning model which can be used for such parameter tuning at a later time. In addition a precision tuning would also require a data analysis for the model output which more closely resembles the HADES data analysis as systematic effects may occur.

\begin{figure}[t]
  \centering
  \includegraphics[width=0.5\textwidth]{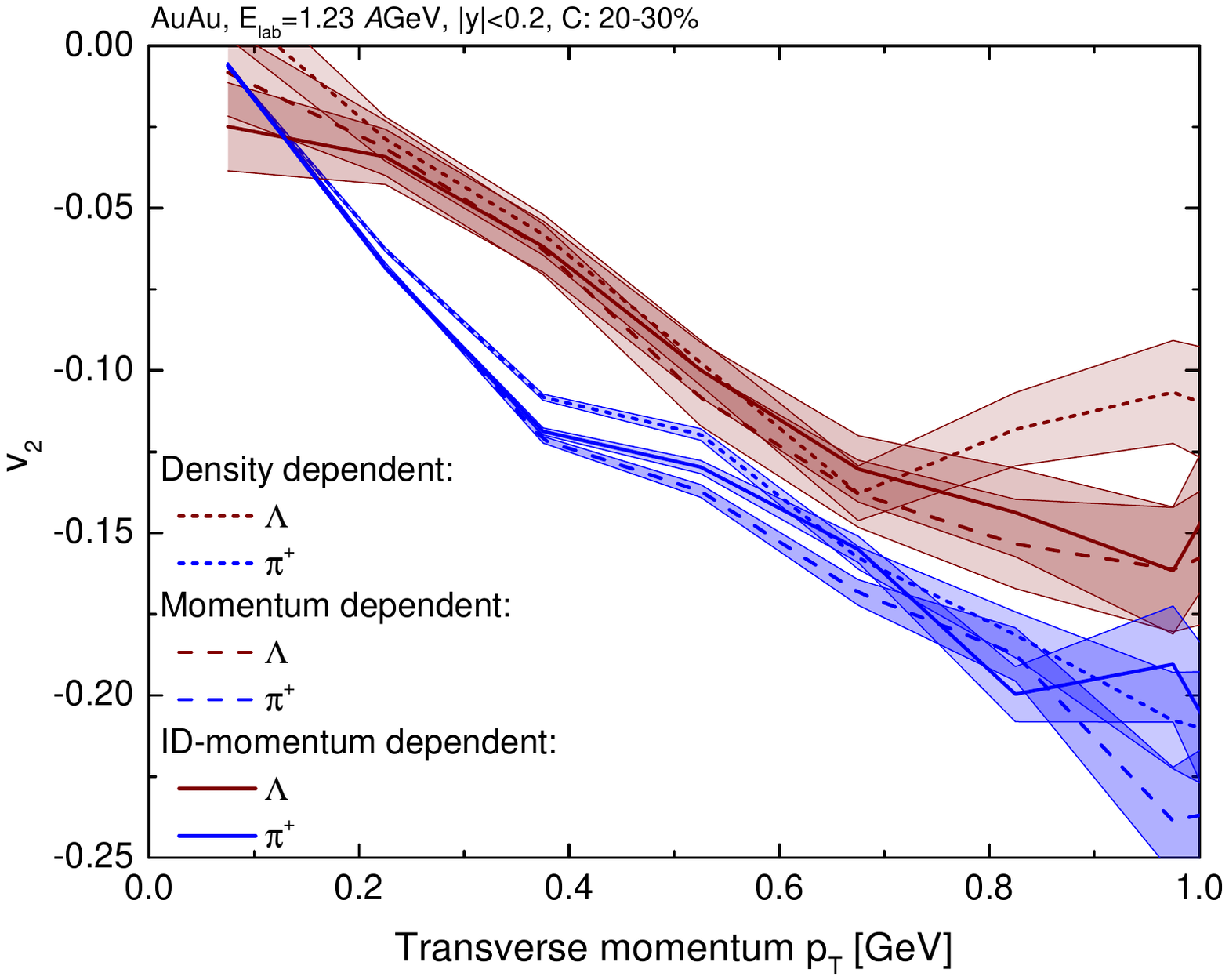}
  \caption{(Color online) Elliptic flow $v_2$ of pions and hyperons at mid-rapidity as function of transverse momentum for Au+Au collisions at $E_{lab}= 1.23 A$ GeV.}
  \label{fig:11}
\end{figure}

As we have seen the proton elliptic flow is not very sensitive on the inclusion of explicit $\Delta$-potentials. One may think that other particles, like the pion or hyperons may be more sensitive to the difference between the momentum dependent and ID-momentum dependent scenarios. Therefore, we present in figure \ref{fig:11} the transverse momentum dependent elliptic flow of pions (blue) and $\Lambda$'s (red, including $\Sigma^0$) for the three different scenarios.

A difference in the elliptic flow for the pions is visible for the two cases without and with momentum dependent potentials, similar to the effect in the proton flow. This means that the pions essentially follow the protons in direction and magnitude of the effect. Again, the difference to the scenario with ID-dependent momentum dependence is small.
%, a small effect at larger transverse momenta is visible. 
The same is true for the hyperon flow which, within the errors bars, does not show any dependence on the potential-implementation.

\begin{figure}[t]
  \centering
  \includegraphics[width=0.5\textwidth]{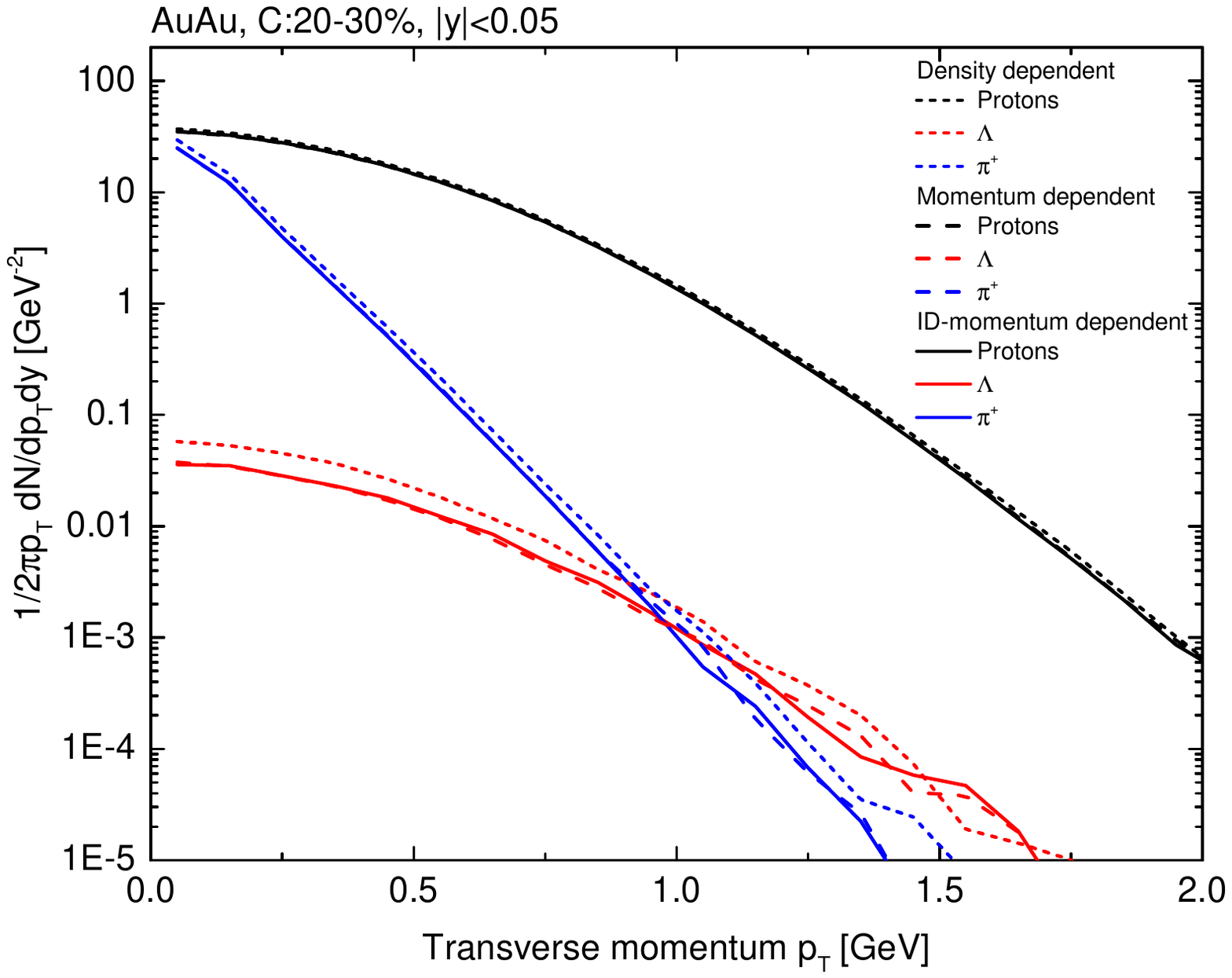}
  \caption{(Color online) Transverse momentum spectra of pions protons and hyperons for Au+Au collisions at $E_{lab}= 1.23 A$ GeV.}
  \label{fig:12}
\end{figure}

Next, we turn to the investigation of the actual transverse momentum spectra of these three hadron species. Figure \ref{fig:12} shows the transverse momentum spectra of protons (black) positive pions (blue) and $\Lambda$'s (blue) for mid-central Au+Au collisions. While the proton spectra stay essentially unchanged some difference can be observed for the pions and hyperons.

\begin{figure}[t]
  \centering
  \includegraphics[width=0.5\textwidth]{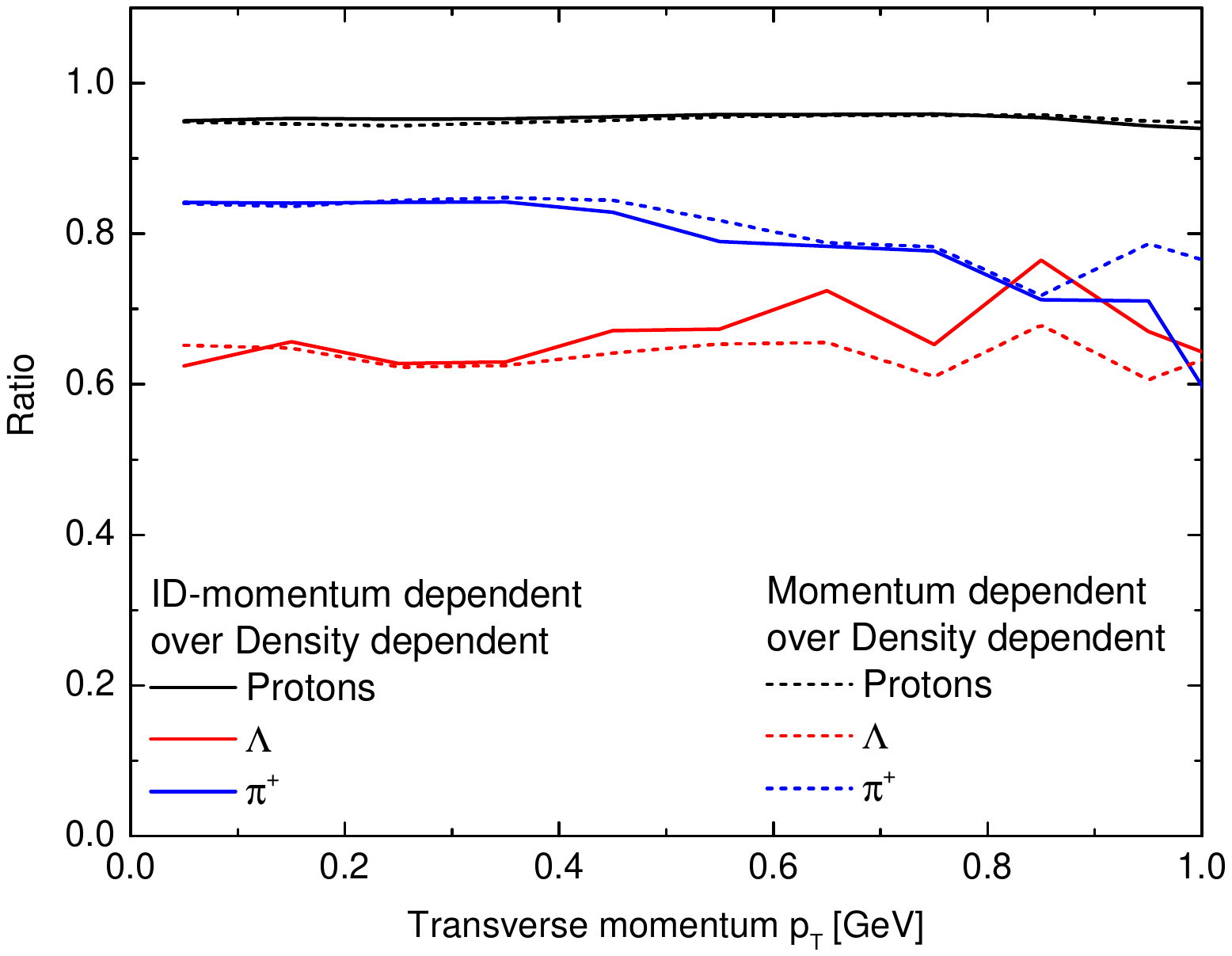}
  \caption{(Color online) Ratio of the transverse momentum spectra of pions protons and hyperons with over without momentum dependent potentials. Shown are UrQMD results for Au+Au collisions at $E_{lab}= 1.23 A$ GeV.}
  \label{fig:13}
\end{figure}

To make the differences better visible, figure \ref{fig:13} shows the ratios of the $p_T$-spectra of the three hadron species with respect to the density dependent scenario. As one can see the shape of the proton spectra does not depend on which type of momentum dependent potential implementation is used. The pion spectra appear to be reduced by a constant factor and a slight softening of the spectrum at large $p_T$ is observed. The hyperon spectrum, however is modified by a significant constant factor, and their spectrum seems to be slightly stiffer if an ID-momentum dependent potential is used. The biggest effect of the momentum dependence seems to be therefore an overall reduction of the mid-rapidity yield of the different particles, where those particle  closer, or even below their elementary threshold energy are reduced more strongly.

\begin{figure}[t]
  \centering
  \includegraphics[width=0.5\textwidth]{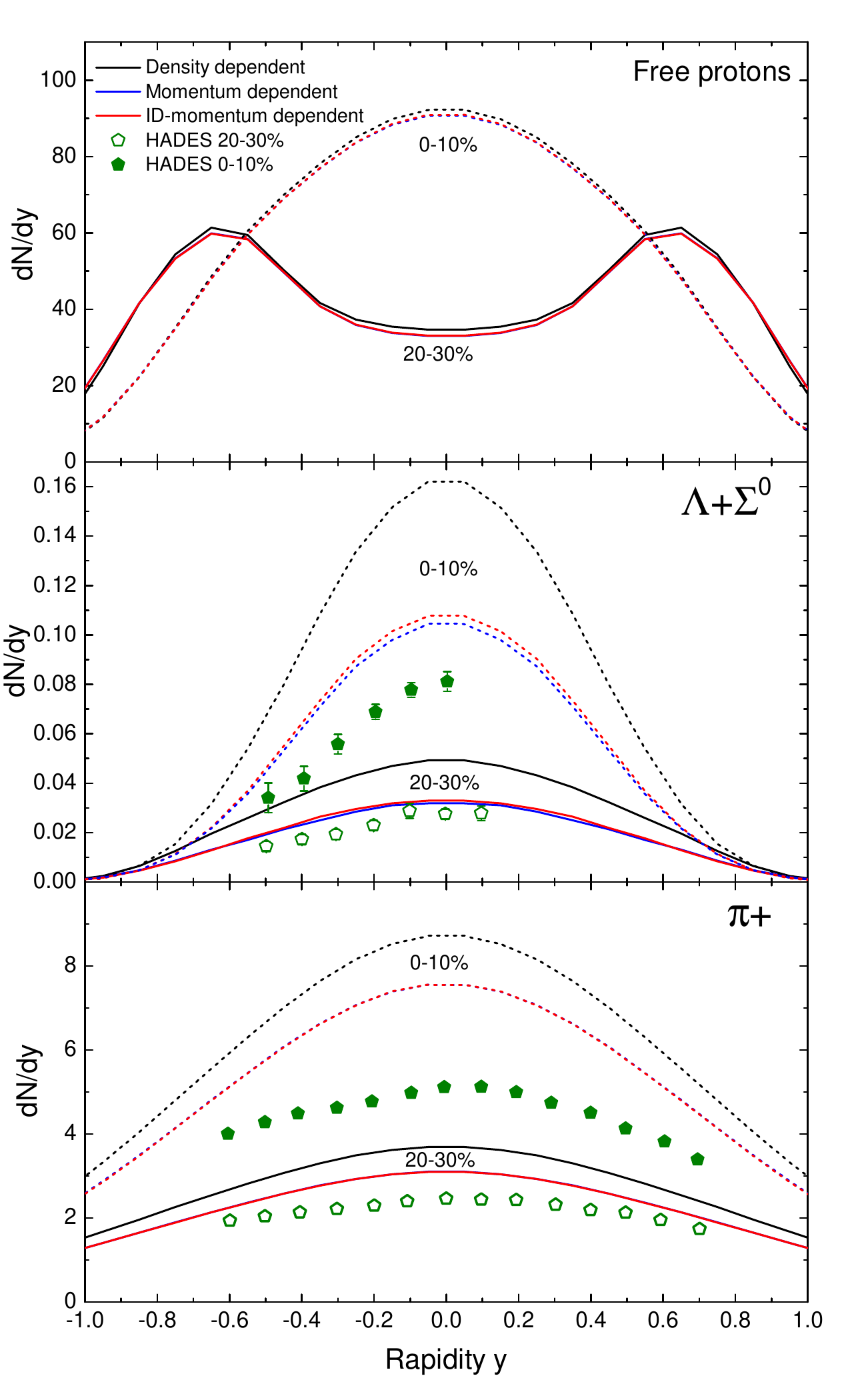}
  \caption{(Color online) Rapidity spectra of protons in two different centrality bins. Shown are UrQMD results for Au+Au collisions at $E_{lab}= 1.23 A$ GeV. Dashed lines correspond to central collisions and solid lines to mid-central. HADES data \cite{HADES:2020ver,tscheib,HADES:2018noy} are shown as green symbols.}
  \label{fig:14}
\end{figure}

To see this effect more clearly, we will finally compare the rapidity distributions of these three hadron species with available data from HADES \cite{HADES:2020ver} in two centrality bins. The three panels in figure \ref{fig:14} show the $p_T$-integrated rapidity distributions of protons, $\Lambda$'s and pions for the three scenarios discussed.

Two different centrality bins for Au+Au collisions at $E_{lab}= 1.23A$ GeV are shown, central (dashed lines) and mid-central (solid lines) collisions. The simulations are compared to HADES data where available. As expected, the proton rapidity distribution does not depend on the implementation of momentum dependent potentials, however, the hyperons and pions do. For the $\Lambda$'s (including the $\Sigma$'s) \cite{tscheib} especially a significant reduction of the yield is observed when momentum dependent potentials are used. For both centralities the inclusion of the momentum dependence leads to a significant improvement of the description of the data. For the pions also a reduction is observed, however, not as significant as for the hyperons which are produced below their elementary threshold. Nevertheless, the average pion per participant, in the scenario with momentum dependent potentials, is $\langle \pi \rangle /N_{\mathrm{part}}=0.175$, which is larger than the HADES result but consistent with the FOPI data at a similar energy \cite{HADES:2020ver}.

Such a suppression of hadron production near the threshold due to momentum dependent potentials has been observed before in simulations with a QMD~\cite{Aichelin:1987ti} and a BUU model \cite{Hong:2013yva}, however the explanation in the QMD is not so straight forward, especially for the pions. We have found that the actual reason for this suppression is a slight reduction of the average center of mass energy of inelastic binary baryon-baryon scatterings which slightly reduces the $\Delta$ generation but does not impact the $\Delta$ absorption as much. This small difference leads to a net suppression of the final pion yield. For the strange hadrons this effect is much stronger as a small reduction of the invariant mass near the threshold energy will have significant effects.

\section{Discussion}
It was shown how the density and momentum dependent single particle potentials from a SU(3) parity-doubling chiral mean field model can be implemented in the non-relativistic version of QMD, more specifically in the UrQMD model. This means we can now consistently implement the momentum dependence of all baryon potentials at every density from a given set of parameters of the CMF model in UrQMD. The effect of the momentum dependent potentials was studied in heavy ion reactions at beam energies from $E_{lab}=0.4 - 30 A$ GeV. 
The general effect on hadron directed and elliptic flow was studied and the effects are qualitatively and quantitatively in agreement with previous studies. The current parametrization of the CMF model provides a moderate momentum dependence and thus the effects are not as strong as in previous simulations where the momentum dependence was implemented in a less consistent way. 

Although the $\Delta$ baryon has a potential different from that of the proton, $U_{\Delta}(p=0,n_B=n_0)=-83$ MeV in accordance with results in \cite{Lopes:2022vjx}, the effect of this difference is negligible for the pion flow \footnote{Note that the effect may be stronger if one assumes a much stronger $\Delta$ potential. See e.g. \cite{Ikeno:2023cyc} and references therein for an in-depth discussion on the $\Delta$ potential.}. Similarly, the effect of the hyperon potentials is very small. 

On the other hand, a significant effect on the hadron multiplicities was observed where a strong reduction of $\Lambda$'s and a moderate reduction of pions in Au+Au collisions at $E_{lab}= 1.23 A$ GeV is observed which significantly improves the overall description of the HADES data. A discrepancy between the UrQMD model and the published pion multiplicities of the HADES collaboration remains, however the pion multiplicity (per participant) with momentum dependent potentials seems to be consistent with previous FOPI data.

Most importantly this work sets the groundwork for several future studies. In this current setup, we can now study experimental observables in the SIS18/SIS100 energy range and their dependencies on the various baryonic couplings in the CMF model. In other words we can use the parameters of CMF as input to study the interactions between hadrons as well as effects of chiral symmetry restoration in the hadronic sector within a consistent description. Furthermore, by changing the parameters of the CMF model, we will be able to implement a first order phase transition with momentum dependent potentials. This will open the route for a direct comparisons of the resulting EoS, extracted from heavy ion collision data, with the neutron star EoS which can be calculated with the same parameters in the CMF, allowing for a direct and easy bridging between these two domains of dense QCD matter studies.

\begin{acknowledgments}
The authors thank J. Aichelin, V. Koch and B. Kardan for valuable discussions about the momentum dependent potentials and HADES data.
T.R. acknowledges support through the Main-Campus-Doctus fellowship provided by the Stiftung Polytechnische Gesellschaft Frankfurt am Main (SPTG). 
T.R. and J.S. thank the Samson AG for their support.
The authors acknowledge for the support the European Union's Horizon 2020 research and innovation program under grant agreement No 824093 (STRONG-2020). This research has received funding support from the NSRF via the Program Management Unit for Human Resources and Institutional Development, Research and Innovation [Grant No. B16F640076]. This work was supported by the DAAD (PPP Thailand).
The computational resources for this project were provided by the Center for Scientific Computing of the GU Frankfurt and the Goethe-HLR.
\end{acknowledgments}

\bibliography{main}% Produces the bibliography via BibTeX.

\end{document}